\documentclass[manuscript,screen]{acmart}

\AtBeginDocument{%
  \providecommand\BibTeX{{%
    \normalfont B\kern-0.5em{\scshape i\kern-0.25em b}\kern-0.8em\TeX}}}

\usepackage{wrapfig}
\usepackage{float}

\usepackage{graphicx}
\usepackage{tikz}
\usepackage{color}
\usepackage{subcaption}

\usepackage{acronym}
\acrodef{PPE}[PPE]{Personal Protective Equipment}
\acrodef{FDA}[FDA]{Food and Drug Administration}
\acrodef{NIH}[NIH]{National Institute of Health}
\acrodef{CDC}[CDC]{Center for Disease Control}
\acrodef{NIOSH}[NIOSH]{National Institute for Occupational Safety and Health}
\acrodef{VHA}[VHA]{Veteran's Health Administration}
\acrodef{DoD}[DoD]{Department of Defense}

\newcommand{\ppe}{\ac{PPE}}
\newcommand{\vha}{\ac{VHA}}

\newcommand{\fda}{\ac{FDA}}
\newcommand{\nih}{\ac{NIH}}
\newcommand{\cdc}{\ac{CDC}}

\newcommand{\etal}{\textit{et al.\ }}
\newcommand{\etals}{\textit{et al's\ }}

\newcommand{\nihEx}{\nih\ 3D Print Exchange}

\newcommand{\clinicalratingperc}{5.6}
\newcommand{\communityratingperc}{4.8}
\newcommand{\notratedperc}{81.3}
\newcommand{\clinicalorcommunityperc}{10.4}
\newcommand{\studyenddate}{January 1st, 2021}
\newcommand{\oursetperc}{83.5}
\newcommand{\numothersub}{103}
\newcommand{\numourset}{520}
\newcommand{\totalsubs}{623}
\newcommand{\numfaceshield}{263}
\newcommand{\percfaceshieldtotal}{42.2}

\newcommand{\nummask}{177}
\newcommand{\percmasktotal}{28.4}

\newcommand{\numstrap}{80}
\newcommand{\percstraptotal}{12.8}

\newcommand{\percothertotal}{16.5}
\newcommand{\numthreedprint}{482}
\newcommand{\percthreedprint}{92.7}
\newcommand{\numlasercut}{49}
\newcommand{\perclasercut}{9.4}
\newcommand{\numinjectionmold}{22}
\newcommand{\percinjectionmold}{4.2}
\newcommand{\numcnc}{14}
\newcommand{\perccnc}{2.7}
\newcommand{\numnomethod}{319}
\newcommand{\percnomethod}{61.3}
\newcommand{\nummultimethod}{173}
\newcommand{\percmultimethod}{33.3}
\newcommand{\numholepunch}{63}
\newcommand{\percholepunch}{12.1}

\newcommand{\nummultifil}{147}
\newcommand{\percmultifil}{28.3}
\newcommand{\numusepla}{223}
\newcommand{\percusepla}{42.9}
\newcommand{\numusepet}{34}
\newcommand{\percusepet}{6.5}
\newcommand{\numusepetg}{140}
\newcommand{\percusepetg}{26.9}
\newcommand{\numuseabs}{64}
\newcommand{\percuseabs}{12.3}
\newcommand{\numusetpu}{47}
\newcommand{\percusetpu}{9}
\newcommand{\numnofila}{332}
\newcommand{\percnofila}{63.8}
\newcommand{\numhadmodeling}{364}
\newcommand{\perchadmodeling}{70}
\newcommand{\numhadslicingsoft}{133}
\newcommand{\perchadslicingsoft}{25.6}
\newcommand{\numuniqueauthors}{448}
\newcommand{\mediandesignsperauthor}{1}
\newcommand{\mindesignsperauthor}{1}
\newcommand{\maxdesignsperauthor}{9}
\newcommand{\numnoaffil}{344}
\newcommand{\percnoaffil}{66.2}
\newcommand{\numproftiaffil}{84}
\newcommand{\percprofitaffil}{16.2}
\newcommand{\numuniaffil}{67}
\newcommand{\percuniaffil}{12.9}
\newcommand{\numhealthaffil}{59}
\newcommand{\perchealthaffil}{11.3}
\newcommand{\numzerocomment}{408}
\newcommand{\perczerocomment}{78.5}
\newcommand{\numonecomment}{56}
\newcommand{\perconecomment}{10.8}
\newcommand{\percatleasttwocomment}{10.8}
\newcommand{\numclinicalrating}{29}
\newcommand{\numcommunityrating}{25}
\newcommand{\numwarningourset}{2}
\newcommand{\percwarningourset}{0.4}
\newcommand{\numwarning}{34}
\newcommand{\percwarning}{5.5}
\newcommand{\numnotrated}{464}
\newcommand{\numnotratedwnotes}{41}
\newcommand{\percnotratedwnotes}{8.8}
\newcommand{\clinicalmask}{2}
\newcommand{\communitymask}{22}
\newcommand{\warningmask}{2}
\newcommand{\notesmask}{28}
\newcommand{\notcheckedmask}{123}
\newcommand{\clinicalfaceshield}{16}
\newcommand{\communityfaceshield}{2}
\newcommand{\warningfaceshield}{0}
\newcommand{\notesfaceshield}{7}
\newcommand{\notcheckedfaceshield}{238}
\newcommand{\clinicalstrap}{11}
\newcommand{\communitystrap}{1}
\newcommand{\warningstrap}{0}
\newcommand{\notesstrap}{6}
\newcommand{\notcheckedstrap}{62}
\newcommand{\numremixes}{179}
\newcommand{\percremixes}{34.4}
\newcommand{\numreplicatedbycomm}{61}
\newcommand{\percreplicatedbycomm}{11.7}
\newcommand{\numpersistentdoc}{183}
\newcommand{\percpersistentdoc}{35.2}
\newcommand{\numhavevideo}{31}
\newcommand{\perchavevideo}{6}
\newcommand{\numlinktorepo}{315}
\newcommand{\perclinktorepo}{60.6}
\newcommand{\numhaveimage}{429}
\newcommand{\perchaveimage}{82.5}
\newcommand{\numpreprocess}{290}
\newcommand{\percpreprocess}{55.8}

\newcommand{\numtested}{44}
\newcommand{\perctested}{8.5}
\newcommand{\numpd}{67}
\newcommand{\percpd}{12.9}

\newcommand{\numccby}{330}
\newcommand{\percccby}{63.5}

\newcommand{\numccbync}{96}
\newcommand{\percccbync}{18.5}
\newcommand{\numccbyncnd}{14}
\newcommand{\percccbyncnd}{2.7}

\newcommand{\perccomplex}{16}
\newcommand{\numspecifiedprinter}{312}
\newcommand{\numcheapprinter}{279}
\newcommand{\perccheapprinter}{89.4}

\newcommand{\numcoverabove}{192}
\newcommand{\perccoverabove}{36.9}
\newcommand{\numNOTfrontofface}{2}
\newcommand{\numcleaning}{84}
\newcommand{\perccleaning}{16.2}
\newcommand{\numcura}{59}
\newcommand{\perccura}{11.3}
\newcommand{\numsimplify}{28}
\newcommand{\percsimplify}{5.4}
\newcommand{\numnoslice}{391}
\newcommand{\percnoslice}{75.2}
\newcommand{\numfusion}{126}
\newcommand{\percfusion}{24.2}
\newcommand{\numsolidworks}{102}
\newcommand{\percsolidworks}{19.6}
\newcommand{\numinventor}{13}
\newcommand{\percinventor}{2.5}
\newcommand{\numrhino}{25}
\newcommand{\percrhino}{4.8}
\newcommand{\numtinkercad}{22}
\newcommand{\perctinkercad}{4.2}
\newcommand{\numnomodel}{161}
\newcommand{\percnomodel}{31}
\usepackage{booktabs}

\begin{document}

\title{Rapid Convergence: The Outcomes of Making PPE during a Healthcare Crisis}


\author{Kelly Mack}
\affiliation{%
  \institution{Paul G. Allen School of Computer Science, University of Washington}
  \city{Seattle}
  \state{WA}
  \country{USA}
}
\author{Megan Hofmann}
\affiliation{%
  \institution{Human Computer Interaction Institute, Carnegie Mellon University}
  \city{Pittsburgh}
  \state{PA}
  \country{USA}
}
\author{Udaya Lakshmi}
\affiliation{%
  \institution{School of Interactive Computing, Georgia Institute of Technology,}
  \city{Atlanta}
  \state{GA}
  \country{USA}
}
\author{Jerry Cao}
\affiliation{%
  \institution{Paul G. Allen School of Computer Science, University of Washington}
  \city{Seattle}
  \state{WA}
  \country{USA}
}
\author{Nayha Auradkar}
\affiliation{%
  \institution{Paul G. Allen School of Computer Science, University of Washington}
  \city{Seattle}
  \state{WA}
  \country{USA}
}
\author{Rosa I. Arriaga}
\affiliation{%
  \institution{School of Interactive Computing, Georgia Institute of Technology,}
  \city{Atlanta}
  \state{GA}
  \country{USA}
}
\author{Scott E. Hudson}
\affiliation{%
  \institution{Human Computer Interaction Institute, Carnegie Mellon University}
  \city{Pittsburgh}
  \state{PA}
  \country{USA}
}
\author{Jennifer Mankoff}
\affiliation{%
  \institution{Paul G. Allen School of Computer Science, University of Washington}
  \city{Seattle}
  \state{WA}
  \country{USA}
}


\renewcommand{\shortauthors}{Mack, et al.}

\begin{abstract}
 The NIH 3D Print Exchange is a public and open source repository for primarily 3D printable medical device designs with contributions from expert-amateur makers, engineers from industry and academia, and clinicians. In response to the COVID-19 pandemic, a collection was formed to foster submissions of low-cost, local manufacture of personal protective equipment (\ppe). We systematically evaluated the \totalsubs\ submissions in this collection to understand: what makers contributed, how they were made, who made them, and key characteristics of their designs. Our analysis reveals an immediate design convergence to derivatives of a few initial designs affiliated with NIH partners (e.g., universities, the Veteran's Health Administration, America Makes) and major for-profit groups (e.g., Prusa). The NIH\ worked to review safe and effective designs but was quickly overloaded by derivative works. We found that the vast majority were never reviewed (\notratedperc\%) while \clinicalorcommunityperc\% of those reviewed were deemed safe for clinical (\clinicalratingperc\%) or community use (\communityratingperc\%). Our work contributes insights into: the outcomes of distributed, community-based, medical making; features the community accepted as ``safe'' making; and how platforms can support regulated maker activities in high-risk domains (e.g., healthcare).
\end{abstract}




\maketitle

\section{Introduction}

Medical making is emerging alongside maker efforts (e.g., hobbyists, engineers, designers, digital fabrication enthusiasts) to apply crafting and digital fabrication to invent, manufacture, and repair medical devices. Research on maker practices across domains has developed rich insights into material practices of collaboration in shared repositories \cite{cheliotisremix_2014, buehler2015sharing, alcock2016barriers} and social norms \cite{fox_2015_hacking, altruism_maker, lindtner_shenzhen}. Unlike many other domains of making, medical making raises vital concerns about safety and efficacy because medical devices can pose significant risks to life and limb. Prior to the COVID-19 pandemic, the  \nihEx\ served as an ``\textit{open, comprehensive, and interactive website for searching, browsing, downloading, and sharing biomedical 3D print files, modeling tutorials, and educational material}'' \cite{nih}---an open repository for medical making. However, as the demand for \ppe\ in the pandemic overwhelmed global supply chains, the \fda\ and \nih\ began sourcing and reviewing alternative, open source designs created by a variety of institutions and hobbyist makers. To ensure that designs are safe to use, the \nih, in partnership with the \vha, \fda, and \cdc, began reviewing these submissions. By analysing this collection, we contribute a better understanding of the effect of a critical, safety-review process on what maker's create, reuse, and share. 

To understand trends from this extraordinary occurrence of medical making, we present a mixed-methods analysis of this \nihEx's COVID-19 Collection. We use a combination of qualitative data from a thematic analysis and quantitative data from web scrapped details of the \totalsubs\ submissions. We reviewed every submission to the COVID special collection between its start date, March 20th 2020, and January 1st, 2021. Our quantitative analysis shows that: submissions rapidly dropped off in April after the initial surge of designs and that most (\oursetperc\%) designs are variations of a few key classes of devices reviewed in this early period. Our qualitative analysis further demonstrates designs within these classes (masks, face shields, and straps) converged to a few common forms.  We also find that makers not affiliated with large institutions struggled to fully document or test their designs. This insufficient documentation of many designs led to wasted \nih\ reviewer time and demonstrates a lack of clarity in maker perceptions of review criteria.

Our results reveal that \nih's goal of collecting diverse and innovative designs from makers was not met. Instead of generating a diverse array of designs, the submission requirements and rating designations led to a rapid convergence of the design space. Even though the \nih\ did not request any particular types of designs, the majority of designs fell into three narrow categories of \ppe: face shields, masks, and straps  (i.e., tension-relief bands, ear-savers). Within these categories, diversity in the designs was low, particularly among face shields where the main defining design feature was the presence or absence of a visor to provide protection from above.

Open maker repositories with no formal review process tend to included a wide range of unique designs \cite{buehler2015sharing, Kuznetsov_2010_expertamateur}. Contrary to these observations, we observed only a few design archetypes and numerous derivatives which made small changes to the manufacturing method (e.g., 3D printer, slicing settings, bed arrangement) or scale (e.g., fit) of the designs. We discuss possible factors for this shortened idea generation phase resulting in a quick design convergence. We further discuss the groups who made these designs, how designers interacted on the forum, and factors that contributed better review outcomes. In short, teams of expert makers outperformed expert-amateur makers in review outcomes. Finally, we discuss how the current review process led to confusion of requirements among makers; this confusion wasted scarce reviewer time.

Based on our findings, we recommend  different ways that maker repositories with review processes can support alternative interactions with the community and yield greater design diversity while maintaining safety. First, make the reviewing process more transparent and effective by 1) ensuring that all key information required for a review is marked mandatory, and 2) providing feedback about why designs received their review rating. Second, introduce a required field to explain updates made to a design in remixes. Small updates can be more rapidly reviewed than more involved changes. Third, pose clear requests to the community. These communications can help ensure that designs diverge rather than converge on what is already positively reviewed. Finally, support and motivate innovation by denoting ``work-in-progress'' submissions and explicitly encouraging designs that diverge from the norm. These features together save reviewers time and position innovation and creativity as values to the community in addition to safety.



Our work lies at the intersection of CSCW themes on material practices summarized in Rosner’s literature review \cite{rosner_matcollab2012}. CSCW scholars have explored material practices in collaborative communities engaged in craft \cite{rosner_matcollab2012}, music \cite{cheliotisremix_2014}, and sites of peer-production \cite{Yamauchi_collableanOSS_2000, Kriplean_articulation2008}. Our work informs and draws from three related themes. The first has developed an understanding of materials and tools as non-human actors in collaborative practices. The second discovers emergent, temporal interactions while the third has evolved around designers' notions of affordances in digital-physical materials. Our inquiry specifically extends Rosner’s observations on the temporal and heterogeneous facets of collaboration with materials \cite{rosner_matcollab2012} aligned with other studies on making \cite{oehlberg2015patterns}. Medical makers balanced safety, urgency, and uncertainty in the digital-physical fabrication process during a pandemic. Their open source contributions required greater articulation work that was critical for design remix and reuse designs in a flexible manner. The repository as a material actor is designed for innovation, not speed, among other factors. We discuss how fabrication work is embedded in material practices of medical making. We provide novel implications for remix, reuse, and coordination to build open source infrastructure for medical making. We reflect on social relations in a community marked by norms of safety, reliability, and regulation. Our findings on rapid design convergence in the \nihEx\ challenge expectations of novelty, variety, and wider reach of open source activity.

\section{Related Work}
\subsection{Digital Fabrication and Peer Production in Medical Making}

Maker activities are characterized by their community’s norms and material practices. Tanenbaum \etal\ describe hobbyist makers tend towards a hedonistic preference of maker-technologies (i.e.,\ 3D printers) that offer speed, replicability, and collective skill to democratize material-driven innovation \cite{Tanenbaum_2013_Democratizing}. In contrast to this technocentric/utilitarian view, others call attention to an ecosystem of sociopolitical actors \cite{lindtner_shenzhen}, community structures \cite{fox_2015_hacking}), and future opportunities \cite{Lindtner_2014_emerginghci} to critique notions of empowerment within material constraints at sites of making. Studies on digital fabrication in healthcare communities reveal how care for recipients of Assistive Technology devices \cite{buehler2015sharing, parryhill2017understanding} and motivations in DIY Health \cite{okane_advances_2016, slegers_makingHealthcare} impact peer production. Relatively less is known about a similar trend in digital fabrication practices applied to medical practice in HCI. 

Medical applications for digital fabrication are on the rise with advances in 3D printing \cite{mukhopadhyay_review_2018, ventola_medical_2014, kalaskar_3d_2017}. Most studies track clinician experiments with the novel use of fabrication technologies in bio-printing \cite{vijayavenkataraman_3d_2016}, surgical guides \cite{malik_three-dimensional_2015}, dentistry \cite{dawood_3d_2015}, implants \cite{curodeau_design_2000}, prosthetics \cite{gretsch_development_2016}, and orthotics \cite{chimento_3d_2011}. This trend correlates to a history of crafting practices and device improvisations \cite{gomez} and open source infrastructures. However, recent HCI studies indicate a wider variety of \textit{''medical makers''} \cite{lakshmi2019point} engage in medical device development and deployment in care delivery roles. They adapt the fabrication process to suit specialized practice \cite{hofmann2019occupational} and generalized care norms \cite {lakshmi2019point}. Hofmann \etal found that occupational therapists limit material iterations to integrate digital fabrication into their standard practices, packed schedules, and keep costs to a minimum \cite{hofmann2019occupational, slegers_makingHealthcare}. Lakshmi \etal discuss how clinician-makers hesitate to distribute prototype designs without due regulatory approval or licensing extending from an ethos of safety and a risk aversion to personal medical liability \cite{lakshmi2019point}. While 3D printing advocates in medicine proved prescient in the COVID-19 \ppe\ crisis \cite{Novak2020_covidremix, lakshmi_covid, hofmann_onlineMakers}, regulatory and policy infrastructure in the medical space is underdeveloped. Similar to the open source software development communities \cite{Yamauchi_collableanOSS_2000}, flexible and ad hoc coordination is key for efficient medical maker response. Medical makers already defer to their professional norms to uphold safety and reliability with risk-averse approaches. In uncertain times, these factors may conflict with expectations of novelty and variety with 3D printing overtly recognized as a tool for innovation within the medical research community (e.g., VA’s Innovator's Network \cite{razak_vha_2018}). It is unclear how these social and material constraints influence peer production mechanisms for medical makers engaged in digital fabrication.

\subsection{Barriers to Reuse and Remix in Digital Fabrication}

Reuse and adaptation of shared designs is a perceived benefit in maker communities \cite{cheliotisremix_2014}. These activities, described as remixing, are motivated by collective learning among makers by contributing to peer production activity on repositories \cite{Kuznetsov_2010_expertamateur, buehler_investigating_2016}. Makers, medical makers included, expect to adapt designs and re-share them with an articulation of their efforts for future reuse. However, these expectations are constrained by factors specific to the digital-physical material process for both adaptation and articulation work acting as barriers to collaboration. 

Unlike physical artifacts, novelty of an adapted digital artifact can be attributed to the extent of variation from the original as Cheliotis \etal note in their study on a musician community \cite{cheliotisremix_2014}. On Thingiverse, Kim \etal describe how popular contributions of preferred digital file types rely on real world constraints around printer filaments and reliable outcomes \cite{Kim2017understanding}. To support collaboration between users, it is better to share the source files generated on modeling tools (e.g., OpenCAD) to retain the original geometry of the model and make editing easier \cite{hudson2016understanding}. However, Alock \etal reports an overwhelming preference for STLs (84\%) over OpenSCAD files (3.7\%) on Thingiverse \cite{alcock2016barriers} possibly because it signals a convenience to download-and-print the model. Regardless of their popularity, STLs are inflexible file types lacking metadata, forcing makers to rebuild designs from scratch. 

When makers share editable models, they fail to articulate key details such as the design’s purpose and manufacturing details (e.g., slicing instructions). Even more experienced users can struggle with inferring the details of a print when there is insufficient documentation on materials, print settings, and/or assembly \cite{Ludwig2014towards, oehlberg2015patterns}. One part of this challenge is that makers rarely document these aspects of their designs as they go, and they avoid the work when sharing online \cite{ashbrook2016towards}. Further, many novices in 3D-modeling struggle to understand the intricacies of models such as dealing with print uncertainties \cite{Kim2017understanding} and figuring how 3D-models interact with real-world geometries \cite{ashbrook2016towards, hofmann2018greater, chen2015encore}. This makes it difficult for them to explicitly document how their design works. One solution may be to integrate the documentation process directly into 3D modeling processes \cite{hofmann2018greater}, however no widely adopted standard tools support this workflow. In the rare case that all relevant information included, variations in printer and filament can still cause prints to fail \cite{Kim2017understanding}. On sharing platforms, insufficient documentation is partly addressed on user forums by the community's discussion on specific 3D-models.This reactive process is not sustainable over time as users continue to remix the model. Documentation can be lost with each iteration leaving gaps for the successive author might not understand everything about the model and be unable to answer questions \cite{flath2017copy, alcock2016barriers}. Moreover, the process increases the burden of articulating their designs on the authors.

Articulation work is embedded in complex cooperative arrangements around the artifact itself \cite{Kriplean_articulation2008, morgan_peerlabor_2014}. For example, on Wikipedia, Kriplean \etals case study analyses how moderators' contribution from core editing shifts to ``meta-work activities'' that ultimately build the collective reputation by overseeing participation, support, and quality of outcome. Morgan \etal in their analysis of alternate WikiProjects found open collaborations persist when they maintain low barriers for participation and community-adapted social structures \cite{morgan_peerlabor_2014}. Most maker communities favor a flexible, informal structure \cite{khanapour2017framing} to maximize participation especially from volunteers \cite{Yamauchi_collableanOSS_2000} over defined roles for critical meta-work to ensure quality. Eventually, this leads to inconsistent information on core properties, evaluation methods, or use cases, leaving most digital fabrication repositories riddled with insufficient documentation of design files. It is not surprising that time constrained medical makers avoid adopting open source designs \cite{lakshmi2019point}. Working within institutional infrastructure, their efforts to make medical devices are further subject to available technical expertise, uncertainties around physical materials, and licensing or regulatory mandates to ensure safe use. Yet, repositories like the \nihEx\ and the limited distribution of digital files on hospital sites indicate that medical makers publish their designs for use, reuse, and distribution. We examine the emergent practices around the recent push to make and design \ppe\ \cite{lakshmi_covid, hofmann_onlineMakers} on the \nihEx. Novak and Loy undertake a wider analysis of COVID-19 response efforts in early 2020 \cite{Novak2020_covidremix}. Our study takes a deep dive into the medical maker community on a single platform. 

\section{Background: The NIH 3D Print Exchange and COVID-19}

The \nihEx\ is a 3D model repository hosted by the US government with an exclusive focus on collecting ``bioscientific'' models. 
Before the COVID-19 pandemic, the collection was an open library of bio-medical models (e.g., molecules, organs), a small collection of open source prosthetic-like devices from e-NABLE, and simple 3D printable lab equipment (e.g., test tube holders). The goal of the project was to be the authoritative source for medical makers' designs. It included the extensive documentation needed to reliably make such models wherever physical equipment was available nationwide. By early 2020, there were efforts to add an expert review process to the exchange that would enable makers to receive feedback from \vha\ and \fda\ experts. The program roll out was hastened to completion in response to the surging COVID-19 pandemic in March. 

Due to the pandemic, traditional \ppe\ and medical device manufactures could not keep up with demand and there were global shortages of respirators, face shields, masks, and other critical supplies. The \nihEx\ released its new review process early through the new COVID-collection. The collections' goal was specifically: ``\textit{to inform decision-making on \ppe\ and medical device production, without stifling innovation...by filtering designs through a systematic review process.''} The collection was intended to connect innovative makers and manufacturers to produce products to fill in supply gaps. Anyone could submit their designs to the collection, queuing them for review by medical and engineering experts within the \nih\ and other government affiliates. 

Makers submit their designs through an extensive, publicly available form \cite{nih_form}. They could provide: a textual description, manufacturing details (e.g., 3D printer model; materials; design files, pre-processing, assembly, cleaning, and use instructions), licensing information, and documentation (e.g., images, testing procedures and data); though, few of these fields were mandatory. All submissions are marked as ``prototypes'' before they are reviewed. Submissions are reviewed based on a priority determined by (1) demand (i.e.,  the design meets an unmet need), (2) feasibility (i.e.,  it seems reasonable that the design works as described), and (3) detail (i.e., the submission includes enough information to make review possible) \cite{nih_faqs}. Reviewed designs are independently produced and tested with actual materials by reviewers to determine what classification, if any, is appropriate. More detailed criteria for review is listed on the collection's FAQ and in a document detailing the different types of masks (general use face masks, surgical face masks, and N95 respirators). Specific criteria for other types of \ppe\ are not present. 

Besides the default prototype status, submissions' review status can be: ``\textit{reviewed for clinical use}'', ``\textit{reviewed for community use}'', and ``\textit{warning}''. Note that none of these terms include the word ``\textit{approved}''; this is a purposeful decision to remove confusion between the exchange's review process and \fda\ approval processes. Positively reviewed designs on the \nihEx\ still do no have \fda\ approval. Submissions reviewed for clinical use are deemed to be the safest and most effective submissions. These are appropriate to use in a high-risk clinical environment. Community use denotes a lower standard where the device itself is expected to be safe but its efficacy cannot be guaranteed; it will not hurt the user, but it might not protect them either. The warning category was used in rare cases where the design itself is not safe. Usually this was received for high risk designs like ventilator parts. If a design did not meet the clinical or community standards but was not so risky to merit a warning, the reviewers would privately provide feedback and leave the design marked as a prototype. Occasionally, reviewers left public comments before deciding the design's final status. We cannot determine if reviewers left comments in all cases or only in the absence of a private email response. 

This review process was quickly overloaded with a surge of new designs. On July 24th, the Exchange stopped considering common face shield and ear-saver designs for review ``\textit{due to the volume of submissions, unless the face shield is a novel design adapted for a specific use}''. They turned their review efforts exclusively to nasal swabs for COVID-19 tests which make up only seven of the \totalsubs designs. 

The \nihEx\ presents an unique opportunity for researchers to study what medical makers do when collectively tasked to address one global problem (i.e., \ppe\ production in a pandemic). Unlike open maker repositories, the \nihEx\ includes an explicit review process and heightened community standards that are in line with the standards clinicians strive to uphold. However, similar to other traditional repositories, makers contributing to the exchange likely still face challenges in learning how to make safely and documenting the quality and safety of their designs so that others can reproduce and build on them.

\section{Methods}
To understand \nihEx\ \ppe\ submissions, we collected fields from each submission for quantitative analysis. We further qualitatively coded \numourset\ submissions made before \studyenddate. We coded three types of \ppe\ which made up the majority of the submissions (\oursetperc\%): face shields, masks, and ear-savers. Each submission was reviewed manually to determine if it was a face shield, mask, ear-saver, or another device. Our final sample of masks, face shields, and ear-savers was \numourset\ of the \totalsubs\ total submissions made prior to \studyenddate.

Based on the submission form structure, for each submission we programatically collected, where ever appropriate, the: 
\begin{itemize}
    \item Entry name
    \item Submission date
    \item Remixing attribution and the original design
    \item Manufacturing method (e.g., 3D printing, laser cutting)
    \item 3D printer model, if applicable
    \item 3D modeling software
    \item Slicing software
    \item 3D printer materials
    \item Review status
    \item External documentation (e.g.,images, videos, PDFs, website links)
    \item Pre- and post-processing instructions
    \item Licenses
    \item Comment counts
\end{itemize}
Each of these pieces of information was either scraped from a well-formatted field on the design submission page or found by searching the text associated with each entry for relevant keywords.

We performed an additional layer of processing on this scraped data to gain insights into makers' reuse of other submissions in their designs (i.e, remixing). The form did not require makers to declare changes made in remixes, though many makers noted it in text. To capture differences across remixes we compared fields between original and derivative designs and logged differences in key fields (e.g., manufacturing method, materials, modeling software, printer-used). Additionally, we searched all text associated with a model for a list of qualifying words that we saw repeatedly in our qualitative coding (e.g., more, less, faster, slower, thicker, thinner, safer).

In addition to this automatically collected data, four authors deductively coded each entry. We derived our codes by inductively coding 50 entries selected through stratified random sampling across the three design categories. Additional codes were developed based on a review of the literature and current media coverage of makers' response to the pandemic. We applied these codes in a top down fashion to all \numourset\ face shield, mask, and ear-saver entries. We all coded in batches of 50 stratified random samples until saturation across the coders was reached, updating and removing codes based on group consensus. We reached saturation with an average inter-rater reliability of 0.87 (range=0.64-1.00) across all accepted codes. Three of these authors went on to individually review the remainder of the data set. We met weekly to update each other and discuss any uncertainties that arose. 

Based on a thematic analysis of these codes, we present themes on the community's values, how trade-offs between values were made in designs, and how remixing behaviors supported convergence of the design space. We developed a shared understanding of the data through weekly meetings where \ppe\ and codes were examined. 

\section{Results}

\begin{figure}
\captionsetup{justification=centering}
\begin{subfigure}{.32\textwidth}
  \centering
  \includegraphics[height=1.4in]{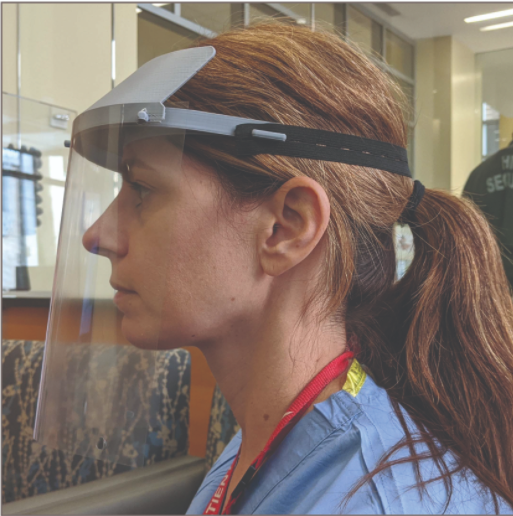}
  \caption{Face shield; \percfaceshieldtotal\% of submissions (3dpx-013359 pictured)}
\end{subfigure}\hfill%
\begin{subfigure}{.32\textwidth}
  \centering
  \includegraphics[height=1.4in]{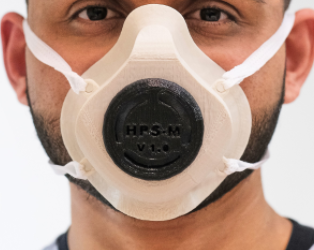}
  \caption{Mask; \percmasktotal\% of submissions (3dpx-013677 pictured)}
\end{subfigure}\hfill%
\begin{subfigure}{.32\textwidth}
  \centering
  \includegraphics[height=1.4in]{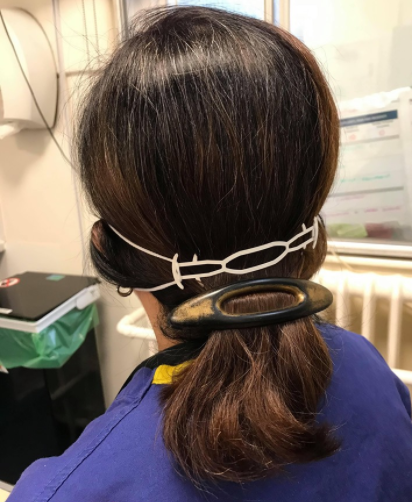}
  \caption{Ear savers; \percstraptotal\% of submissions (3dpx-013860 pictured)}
\end{subfigure}\hfill%
\caption{Examples of the three types of \ppe\ we use in our analysis. (a) shows a face shield (\percfaceshieldtotal\%), (b) shows a mask (\percmasktotal\% of submissions), and (c) shows a tension relief band (ear-savers comprised \percstraptotal\% of submissions).}
    \label{fig:ppe}
    \Description{The figure shows the most popular three types of maker made \ppe.
(A) is a face shield. It has a 3D printed frame that goes around the face, touching the face in the forehead. There is a small part that covers the forehead from above contaminants. Attached to the front of the frame is a clear plastic sheet that protects the face from contaminants from the front.
(B) is a a mask. Like a surgical mask, it covers the mouth and nose of the person. This mask is 3D printed and has a pear-shape. There is a black part attached to the snout (front) of the mask that contains the filter material. It is attached to the face with elastic straps that wrap around the back of the head.
(C) is a tension relief band. It is a small 3D printed bad that is about the size of a bookmark. It has notches at both skinny ends which allow rubber bands/straps to wrap around them. These means that the plastic holds the tension of these straps, rather than the ears.}
\end{figure}

In March 2020, the \nih\ opened the 3D Print Exchange as a place for people to post design ideas and generate discussion and feedback. In our dataset of the \totalsubs\ submissions between March 20th and January 1st, the designs fell into three main categories: face shields (N=\numfaceshield/\totalsubs, \percfaceshieldtotal\%), face masks (N=\nummask/\totalsubs, \percmasktotal\%), and ear-savers (N=\numstrap/\totalsubs, \percstraptotal\%) (\autoref{fig:ppe}). The remaining submissions (N=\numothersub/\totalsubs, \percothertotal\%) included mask cases, ventilator parts, or door-openers. In this section, we characterize the dataset of face shield, mask, and ear-savers that we qualitatively coded (N=\numourset). First, we discuss key properties; how they were designed, manufactured, and by whom. Then, we narrow our focus to key properties for medical making: replicability and safety.

\begin{figure}
\captionsetup{justification=centering}
  \centering
  \includegraphics[width=\textwidth]{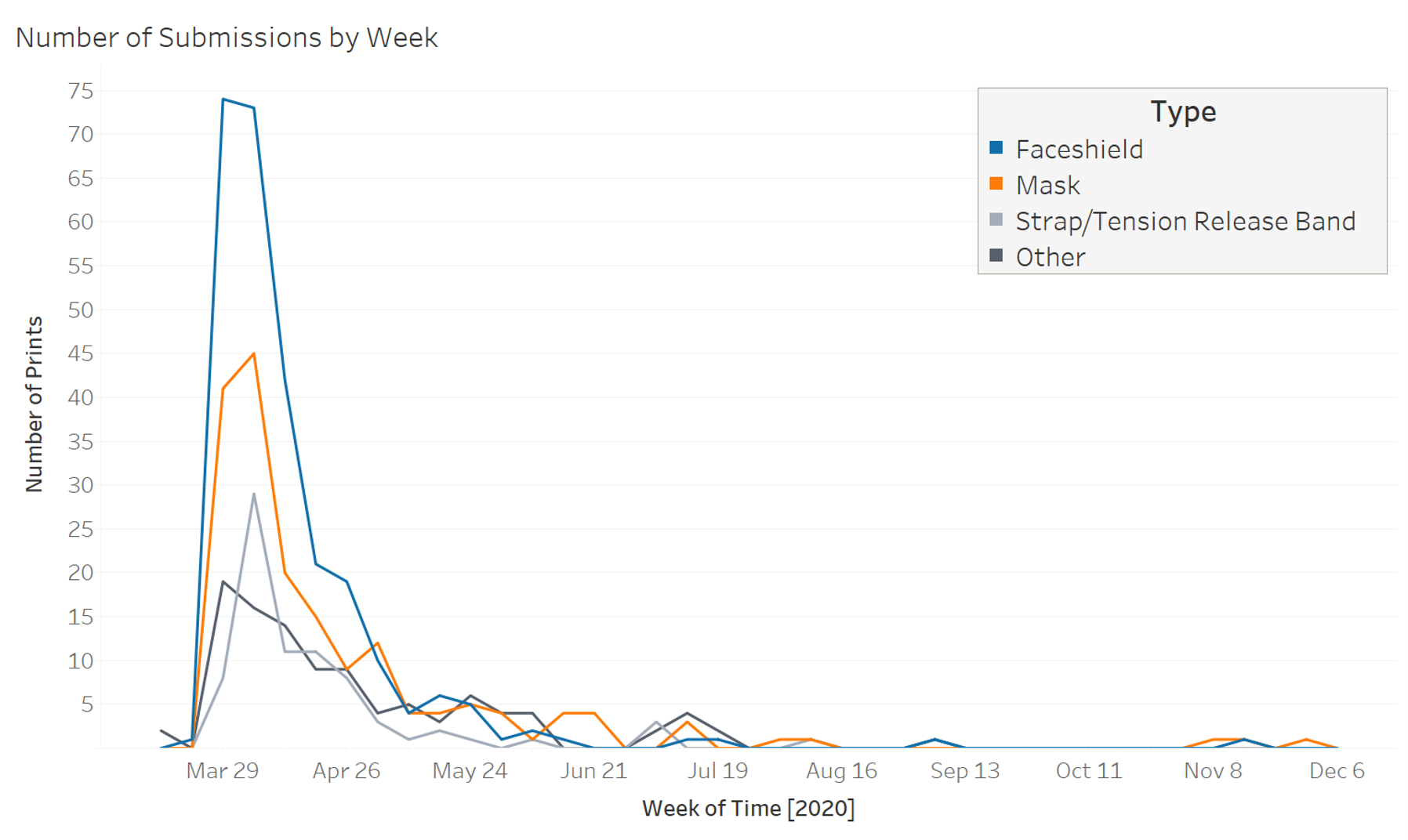}
  \caption{The number of submissions per type of \ppe\ were most popular at the end of March and early April. Face shields are denoted by blue bars, masks by orange, and ear-saver by grey.}
    \label{fig:subs_over_time}
    \Description{ A a bar graph that shows the number of submissions of each type of entry - face shield, mask, and ear-saver - as a function of time. The x-axis is split up into time intervals by week, starting with March 29 and ending December 31. In each time interval, the left most bar represents the number of face shields submitted, the middle bar represents the number of masks submitted, and the right most bar represents the number of ear-savers submitted. We see an increase in number of submissions of all types from March to April and then an exponential decrease from April on wards. }
\end{figure}

\subsection{Temporal Trends}
Submissions surged right after the collection was created in immediate response to the pandemic in the United States. The total number of submissions steadily increased until and peaked in the first week of April (\autoref{fig:subs_over_time})
then the submission rate dramatically decreased. It increased slightly with the resurgence of the virus in the United States in May.

Makers tended to submit designs with greater perceived importance  or complexity. The first submissions (prior to March 29th) were two ventilator valves and one face shield, which are simpler to model and manufacture than a face mask. The media had also expressed that these were more important for saving lives than ear-savers which only increase mask comfort. This is an early example of a repeated pattern in our data. Makers tended to focus on designing what attracts attention or was deemed important, especially by news sources, rather than what could be reliably produced to fulfill \ppe\ needs.

\subsection{Material Trade-offs}

Due to resource scarcity induced by the pandemic, makers made careful trade-offs when selecting manufacturing methods and materials. Makers had to balance between competing goals of broadening participation, using  available materials, rapid manufacturing, and the safety of a design. We present three examples below that highlight these tensions and the trade-offs that were made that were perceivable in the designs. 

\subsubsection{Material Selection, Safety, and Participation}

Many submissions could be made by expert-amateur makers. The most common filaments used were all widely available to consumers: PLA (N=\numusepla/\numourset, \percusepla\%), PET (N=\numusepet/\numourset, \percusepet\%), PETG (N=\numusepetg/\numourset, \percusepetg\%), and ABS (N=\numuseabs/\numourset, \percuseabs\%). PLA, PET, and PETG are common and easy to print with. ABS, however, requires more advanced setups due to toxic off-gassing. 
Similarly, for designs that specified a particular 3D printer, the majority (\numcheapprinter/\numspecifiedprinter, \perccheapprinter\%) used printers available to consumers for less than \$10,000. Many submissions listed multiple filament options (N=\nummultifil/\numourset, \percmultifil\%) (e.g., printing a face shield in PLA or PETG). Notably, the most commonly remixed face shields (3DPX-013306, 3DPX-013359) could be made with several variations of PLA or PETG, and could be manufactured on a consumer printer. The prevalence of easy-to-use materials afforded opportunities for hobbyists and broadened participation.

On the other hand, more complex materials or printers could improve safety at the expense of participation. \perccomplex\% of designs used materials that require special equipment or additional expertise to work with (e.g., TPU, Nylon, PC, ASA). The most commonly remixed mask was printed with nylon that requires an industrial printer. Nylon was chosen because it can be sanitized, unlike PLA or PETG. Therefore, substitutions other filaments could be unsafe. Similarly, some designs combined multiple filaments to meet particular design goals at the expense of easy manufacturability. For example, the ``Helmet-Compatible Community Face Mask'' (3DPX-013354) used a rigid material (e.g, PLA, ABS, PETG) for the snout to ensure the filter was held away from the nose and mouth. It used a flexible material (e.g., TPU) where the mask touches the face to improve comfort and air-seal. Choices by some makers to trade off manufacturability for other goals shows that they believed advanced methods were required at the cost of supporting more makers. 

\subsubsection{Powerful Tools that Limit Participation}

A design's manufacturing method determines who can make a design and how much work is required. Unsurprisingly, 3D printing was by far the most popular method (N=\numthreedprint/\numourset, \percthreedprint\%) followed by laser cutting (N=\numlasercut/\numourset, \perclasercut\%) and injection molding (N=\numinjectionmold/\numourset, \percinjectionmold\%).  The fact that most designs supported 3D printing by hobbyist makers broadened who could manufacture \ppe.

Several entries listed more than one manufacturing method (N=\nummultimethod/\numourset, \percmultimethod\%), such as the ``Georgia Tech Face Shield for Injection Molding, 3D Printing, Waterjet, Laser Cutting'' (3DPX-013314). Often these gave makers choices. For example, the ``NAVAIR - TDP for 3DVerkstan Protective Face Shield'' (3DPX-014090) lists that the submission can either be ``\textit{printed on non-industrial 3D printers or laser cut.}''  While 3D printers are relatively slow and require post processing, they are widely available. Injection moldering, on the other hand, is fast but inaccessible to most for hobbyists. Makers designed for multiple manufacturing methods to both support makers and increase manufacturing efficiency. 
 
Other designs utilized multiple manufacturing techniques for the same design. For example, the ``Southern Tier Face Shield'' with model ID 3DPX-014082 was one of several face shield designs that required a 3D printed frame that goes across the wearer's forehead and a laser cut PC barrier to prevent droplets from reaching the face. They chose materials like PC because they can be quickly and automatically cut. Alternatively makers may increase post processing requirements to avoid using additional manufacturing machines. For example, regular, office hole-punchers could be easily used with transparent, plastic, 3-ring binder sheets to create the clear face shield without laser cut plastic (N=\numholepunch/\numourset, \percholepunch\%). The ``Livingston Shield v2.2'' (3DPX-014416) instructs users to use a hole-puncher to create 4 holes in a transparency sheet to attach to the 3D printed face shield frame. Though the materials were common and unlikely to run out in the pandemic, this design requires more manual post processing to punch and attach the sheets to the 3D printed frame than laser-cut alternatives. Makers traded-off increases in production speed through advanced manufacturing with slow manual process that increased participation.

Aside from manufacturing tools, makers' choices of software can effect participation. The most used modeling and slicing tools are listed in \autoref{tab:manufacturingData}. Most modeling tools were oriented towards professionals (e.g., Solidworks) or expert-amateurs (e.g., Autodesk Fusion 360). These require more experience to use than novice-oriented tools (e.g., TinkerCAD). \perchadmodeling\% (N=\numhadmodeling/\numourset) and \perchadslicingsoft\% (N=\numhadslicingsoft/\numourset) of submissions provided the modeling and slicing software used, respectively. The ability to edit another user's model depends on the software used (e.g., it is difficult to edit a model made in Blender in Solidworks).

\begin{table}[]
\centering
\caption{Submission Counts and data set percentages for reported manufacturing methods, 3D printer filaments, CAD tools, and Slicing Tools. }
\label{tab:manufacturingData}
\begin{tabular}{ccc}
\toprule
\textbf{Manufacturing Method} & \textbf{Submission Count} & \textbf{Portion of All Submission} \\ \midrule
3D Printing                   & \numthreedprint                       & \percthreedprint\%                             \\
Laser Cutter                  & \numlasercut                        & \perclasercut\%                              \\
Injection Mold                & \numinjectionmold                        & \percinjectionmold\%                              \\
CNC                           & \numcnc                        & \perccnc\%                              \\
None Reported                 & \numnomethod                        & \percnomethod\%                               \\ \midrule
\textbf{3D Printer Filament} &                           &                                    \\ \midrule
PLA                           & \numusepla                       & \percusepla\%                             \\
PET                           & \numusepet                       & \percusepet\%                             \\
PETG                          & \numusepetg                       & \percusepetg\%                             \\
ABS                           & \numuseabs                        & \percuseabs\%                             \\
TPU                           & \numusetpu                        & \percusetpu\%                              \\
None Reported                 & \numnofila                        & \percnofila\%                               \\ \midrule
\textbf{CAD Tool}             &                           &                                    \\ \midrule
Fusion 360                    & \numfusion                       & \percfusion\%                             \\
SolidWorks                    & \numsolidworks                        & \percsolidworks\%                             \\
Autodesk Inventor             & \numinventor                       & \percinventor\%                             \\
Rhino                         & \numrhino                        & \percrhino\%                              \\
TinkerCAD                     & \numtinkercad                        & \perctinkercad\%                              \\
None Reported                 & \numnomodel                        & \percnomodel\%                               \\ \midrule
\textbf{Slicing Tool}         &                           &                                    \\ \midrule
Cura                          & \numcura                        & \perccura\%                             \\
Simplify3D                    & \numsimplify                        & \percsimplify\%                              \\
None Reported                 & \numnoslice                        & \percnoslice\%                                 \\ \bottomrule
\end{tabular}
\end{table}

\subsection{Community Members and Interactions}

Prior work positions maker communities as mainly hobbyists working on independent projects in a shared space \cite{khanapour2017framing}. But the \nihEx's community additionally included what we call \textit{affiliated makers} who were affiliated with a university, healthcare facility, and/or for-profit company. Affiliated makers usually represented larger teams of experts. Interaction between community members on the exchange was uncommon, making it difficult for makers to seek support through the repository.

\subsubsection{Individuals and Affiliated Teams}

The \nihEx\ was built to support the open exchange of designs and foster collaboration across stakeholders (e.g., healthcare professionals, universities, companies, entrepreneurs, hobbyist makers). \numuniqueauthors\ unique authors submitted designs. The median number of designs submitted per person was  \mediandesignsperauthor\ and the range was \mindesignsperauthor-\maxdesignsperauthor. Our qualitative review revealed that most (N=\numnoaffil/\numourset, \percnoaffil\%) authors listed no affiliation with their submission. We suspect this indicates a lone maker who is not affiliated with a relevant organization. Those submissions with listed affiliations had team members from industry (N=\numproftiaffil/\numourset, \percprofitaffil\%), academia (N=\numuniaffil/\numourset, \percuniaffil\%), and the healthcare industry (N=\numhealthaffil/\numourset, \perchealthaffil\%).

\begin{figure}
    \centering
    \includegraphics[width=.5\linewidth]{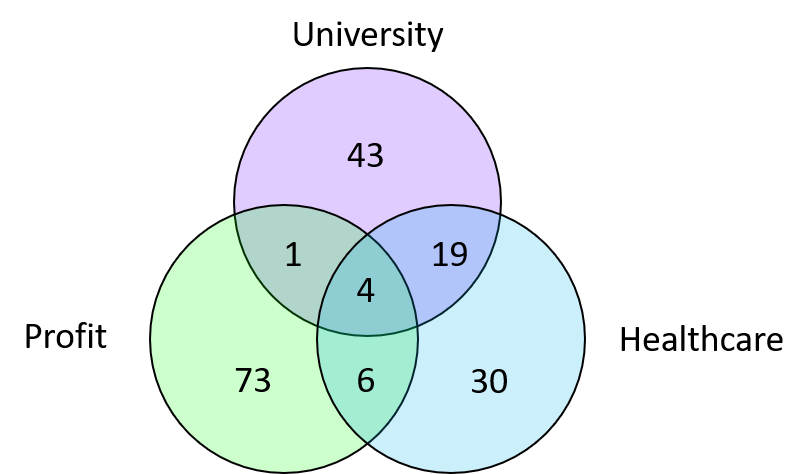}
    \caption{The 176 mask, face shield, and ear-savers that were designed by an affiliated person divided up according to affiliation of the members. Though most designs were carried out by a single type of organization, we see 30 designs with multiple types of contributors.}
    \label{fig:affiliations}
    \Description{A Venn diagram showing the number of designs created by universities, for profit companies, and health care workers. For the designs that only had one type of affiliation, 43 were university, 73 were for-profit companies, and 30 were healthcare. 1 design was by for-profit and university, 6 designs were for-profit and healthcare, 19 designs were university and healthcare. 4 designs had all three of these affiliations.}
\end{figure}

Numerous designs were the result of collaborations within and across institutions. As shown in \autoref{fig:affiliations}, 30 projects involved people with different affiliations. The most common type of collaboration was between universities and health care facilities (N=19/30, 63.3\%). The ``Stopgap Surgical Facemask'' (3DPX-013429) lists 59 team members from for-profit institutions, universities, hospitals, the \fda, and the \vha. 
While affiliated makers often worked in teams, unaffiliated makers rarely collaborated.

\subsubsection{Community Interactions}
The \nihEx\ supports lightweight interaction between makers through submission comments, but commenting was rare. \perczerocomment\% (N=\numzerocomment/\numourset) of designs had zero comments and \perconecomment\% (N=\numonecomment/\numourset) had only one comment. Based on studies on other studies of COVID-19 medical making \cite{hofmann_onlineMakers, lakshmi_covid}, we expect that makers were primarily communicating in other platforms. On the exchange itself, feedback between makers was not the norm. For instance, the ``Helmet-Compatible Community Face Mask'' (3DPX-013354) designer stated in the submission ``\textit{I welcome all feedback in the comments section to further iterate and optimize}'', and the ``USCSW modified 2 part build with Pencil Popper'' (3DPX-013404) mask designer after receiving a 1/5 star community-rating on the \nihEx\ commented ``\textit{Please contact me to explain the one star - I'd be happy to modify anything you didn't like}''. Neither designer received a response. Among the \percatleasttwocomment\% of designs that had at least two comments, constituting a conversation, the median number of comments was 3. The only outlier was the, ``Stopgap Surgical Face Mask (SFM)'' (3DPX-013429), the first revision of a mask that later received clinical usage rating (see 3DPX-014168), that had 111 comments.
We found no evidence of the \nihEx\ broadly being used to collaborate or directly communicate between makers and/or other stakeholders (e.g., \nih, clinicians). 

\subsection{Replicability and Documentation}
For the  \nihEx\ to be useful, makers need to be able replicate submissions. We found evidence of remixing behavior (\numremixes/\numourset, \percremixes\%), but only found  \numreplicatedbycomm\ (\percreplicatedbycomm\%) entries that comments reported as successfully replicated. Thus, remixing was prevalent, but its unclear if they were manufacturing others' designs. We have no way of measuring the number of people who made a design and chose not to share that on this site. Thus, we examine other factors which may influence replicability (e.g., documentation, ease of manufacturing, and licensing). We expect that submissions with more complete documentation, that are easy to make, and have open licences (e.g., public domain) would be more readily adopted. Other factors, such as media attention or affiliation with famous groups (e.g., Prusa, e-NABLE) are also likely contributors beyond the scope of this study. 

\subsubsection{Prototype Remixes}
\begin{figure}
    \centering
    \includegraphics[width=.8\linewidth]{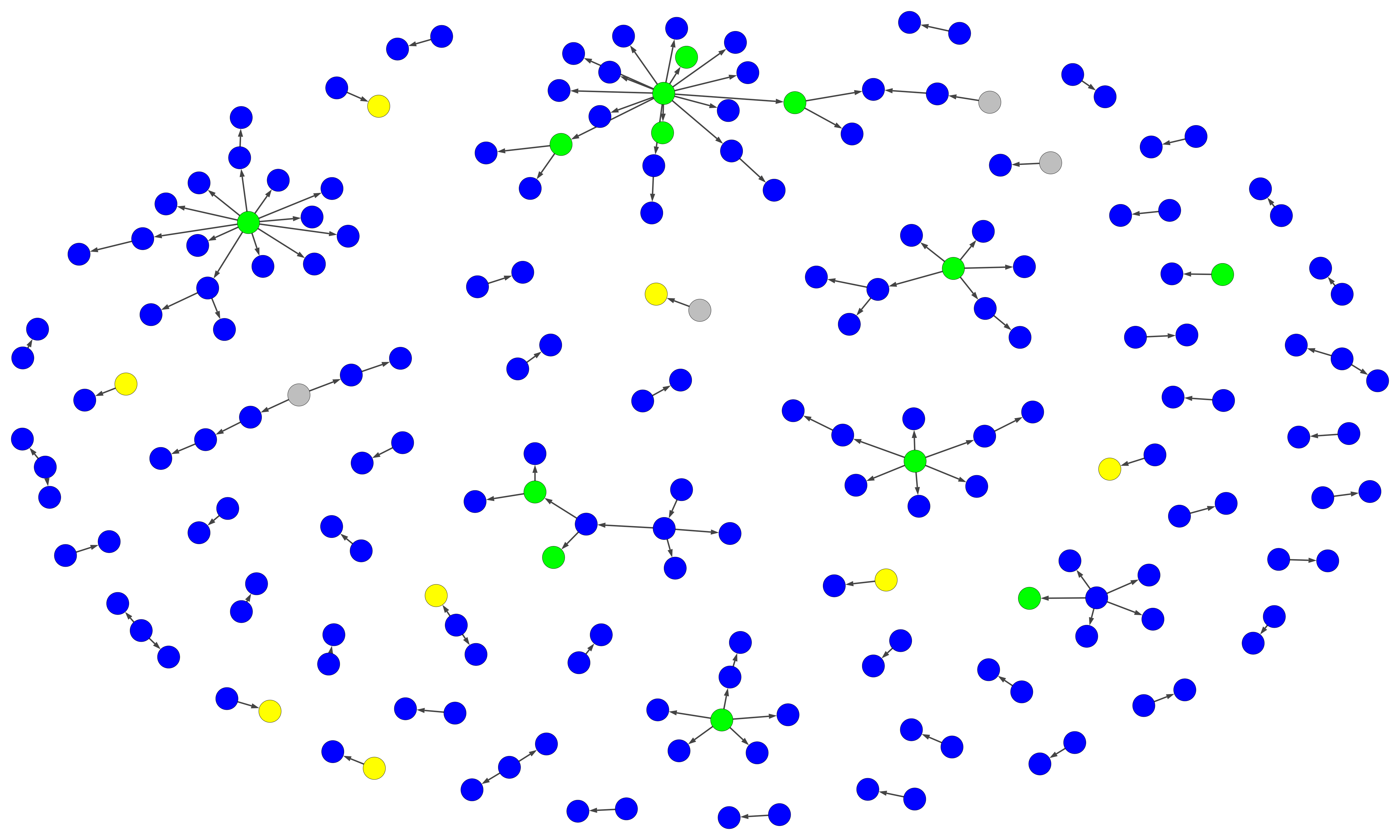}
    \caption{A network showing remixing relationships. An arrow starts at the original design and points to the remix of that design. Colors represent usage rating, with blue nodes as unreviewed, yellow nodes as rated for community usage, and green nodes as rated for clinical usage. The two grey nodes are designs that linked to pages that no longer exist.}
    \label{fig:remixes}
    \Description{A network graph consisting of colored circular nodes and arrows pointing to other nodes. An overwhelming majority of the 186 nodes are blue, meaning unreviewed, and only have one or two edges. There are a few very dense parts of the graphs where there are a lot of edges. Notably, one cluster with 17 edges are all coming from one clinically reviewed design (the DTM face shield). There is another cluster of 12 edges coming from another clinically reviewed design (the Verkestan face shield). There are two smaller clusters for the most popularly remixed strap and mask, each with around 5 edges.}
\end{figure}

The \nihEx\ facilitated collaboration and iteration for ``remixing'', similar to other popular maker forums like Thingiverse and Instructables \cite{cheliotisremix_2014, oehlberg2015patterns}. 131 out of \numourset\ (25.2\%) of entries were listed as remixes or ``other versions'' of models on the \nihEx. \autoref{fig:remixes} presents the remixing network (186 designs), omitting submissions that are neither a remix or remixed. Many nodes (56, 30.1\%) were only remixed once. 
There were notable outliers: one design, the ``3DVerkstan 3D printed face shield head band'' (3DPX-013306), was remixed 12 times and 4 additional designs were derivative of those remixes. Another, the ``DtM-v3.1 Face Shield \ppe'' (3DPX-013359), was remixed 16 times with 6 additional derivatives. Both of these designs were made by affiliated makers. ``3DVerkstan 3D printed face shield head band'' is made by 3DVerkstan, a European 3D printing company, and the ``DtM-v3.1 Face Shield \ppe'' involved team members from Microsoft, three universities, and three hospitals. The mask and ear-saver that had the highest number of remixes were the ``Stopgap Surgical Face Mask'' (3DPX-013429) (5 remixes), which was made by an expansive team crossing companies, universities, and hospitals, and the ``Surgical Mask Tension Release Band for Ear Comfort \& Extended Use'' (3DPX-013410) (6 remixes), which was designed by a \vha\ employee. It is important to note that three of these four designs 
were rated for clinical use, and that no designs in our remixing graph were given a warning usage rating. 
Makers did not iterate to remix designs flagged with warnings to fix those flaws; they remixed successful designs to work under their local manufacturing constraints. Overall, we see that safety and affiliation of designs influenced remixing behavior. This implies that safety was a community norm and affiliated makers were trusted sources of designs.

Some remixing behavior is not captured by explicit links between submissions. For example, many designs shared a similar shape to the popular  ``Montana Mask'' (3DPX-013443) which was spotlighted on Good Morning America on April 12th \cite{goodmorning}. Further, not all makers attributed credit. For example, the maker of the ``3 Hole Punch Minimal Face Shield'' (3DPX-013501) found that someone had remixed their design  by putting two copies of the original design in their printing file without attributing. They commented: ``\textit{at least credit the creator}''. 

Our qualitative analysis showed that remixes were primarily incremental changes to support alternative manufacturing techniques. Few changes were intended to significantly influence use or efficacy. 54 of the remixes listed a change in materials, of which 20 added new materials not mentioned in the original design.
25 designs strictly limited the number of materials options/materials used in a design. However, the majority (N=17) of these designs only removed  complex filament to use (e.g., TPU, Nylon, ABS). 21 remixes used different modeling software than the original submission, which may make it easier for makers to replicate the design in the CAD tool of their choice. 13 remixes used different 3D printers models enabling more people to manufacture the design and 6 tailored a design suited for ``many'' printers for an individual printer model. For example, the ``FDM Printable version of Stop Gap Mask'' (3DPX-013771) remixed the popular ``Stopgap Surgical Mask'' to make it ``\textit{allow printing on hobby style FDM printers (PLA, PET-g etc)}''. The original design required an industrial Powder Bed Fusion Nylon printer. Note that this change effects the mask's porosity, making it harder to to disinfect. Other common reasons for remixing designs included adjusting designs to fit different size print beds, take less time to manufacture, require less material, or to print more than one design at a time. Occasionally, designs affected comfort or ease or use in small ways (e.g., ``[This change] makes it a bit more comfortable for different head sizes'' (3DPX-013659)) While some of these changes may effect safety, none constitute divergence from the original design. On the \nihEx, remixing behavior was almost exclusively tweaking designs to support new makers. 

There are a few examples of substantial feature changes, often motivated by local user feedback. One face shield design, ``Anvil Verkstan Visor'' (3DPX-014089), significantly modified the popular ``3DVerkstan V3 - Face Shield'' based on community feedback: ``\textit{The entire visor has been redesigned and model[ed] from scratch so there will be variances in widths, curves, length, etc. when compared to the original. We re-made this model to better support our local community in our efforts to help the workers on the front lines.}'' Another design, the ``Surgical Mask Tension Release Band with Hair Stabilizer'' (3DPX-013819), iterated on a clinically reviewed design to improve it based on issues experienced by clinician users: ``\textit{They requested a way to keep the band from moving around/flying off while attempting to put on or take off their masks. I incorporated a section of hair pick so that the part can be inserted into the hair, where it will stay on it's own, allowing both hands to be used for putting on or taking off the mask.}'' We observed few remixes like these, which implies that makers either created designs from scratch when addressing more significant design requirements (e.g., clinical usage, fit) or that more makers were interested in tweaking designs to support manufacturing under their resource constraints.

\subsubsection{Documentation}
The \nihEx\ was created for sharing \ppe\ materials for collaboration, education, and practical use. Thus, it is crucial for entries to be documented to foster communication between makers, reviewers, manufacturers, and \ppe\ users. Documentation was often presented as static documents (N=\numpersistentdoc/\numourset, \percpersistentdoc\%) (e.g., PDFs), and video links (N=\numhavevideo/\numourset, \perchavevideo\%).   Images were also a popular form of documentation. All entries included at least one thumbnail image, by default a view of the 3D model, and the majority included additional photographs or diagrams (N=\numhaveimage/\numourset, \perchaveimage\%). A majority of entries included at least one web link (N=\numlinktorepo/\numourset, \perclinktorepo\%), often to a portfolio or alternate repository (e.g., Thingiverse, GitHub). External website content is dynamic, but the \nih\ required static documentation to be included on the exchange itself. We stopped reviewing links because we found several broken links during our qualitative analysis. Finally, a majority of entries (N=\numpreprocess/\numourset, \percpreprocess\%) also included pre/post processing information, such as printer settings, cleaning instructions, and material recommendations, which are critical to ensure proper manufacturing and safe use. 
Overall, makers tended to provide documentation that required the least additional work from them, preferring easy to update websites over creating static documents, or adding easy to capture images instead of videos.
Documentation did not appear to be makers' top priority.

In the medical domain, reproducible testing procedures and results are critical. Test results are needed to quantify the level of protection a design provides. Their importance to reviewers is supported by the correlation between presence of testing and community or clinical approval ($\chi^2 = 4.1, p < .05$). Only \numtested\ (\perctested\%) designs documented rigorous testing results: 6 face shields and 46 mask. A $\chi^2$ test reveals that affiliation with healthcare facilities or universities correlated with the presence of testing results (healthcare: $\chi^2 = 22.3, p < .00001$ ; university: $\chi^2 = 21.4, p < .00001$). This is likely because testing requires specialized equipment that consumers cannot easily access. The community's importance of testing advantage affiliated makers over unaffiliated makers.

\subsubsection{Licensing}
All submissions included a licenses that set permissions for sharing, adapting, selling, or remixing submissions. Several designs offered little or no restrictions to use and adaptations: N=\numpd/\numourset, \percpd\% were public domain and N=\numccby/\numourset, \percccby\% were CC BY or CC BY-SA; both of these do not restrict usage, but require attribution to the original. Other designs limited themselves to non-commercial use only (N=\numccbync/\numourset, \percccbync\% were CC BY-NC or CC BY-NC-SA), while still others used the strictest of licenses which do not allow for modifications to be made to the design (N=\numccbyncnd/\numourset, \percccbyncnd\% were CC BY-NC-ND or CC BY-ND). Overall, the tendency for authors to use less restrictive licences is aligns with prior work that shows that maker communities tend to value openness \cite{Kuznetsov_2010_expertamateur}. 

\subsection{Convergence of Designs}
Our dataset was characterized by rapid convergence of design ideas; there was little exploration of new forms of \ppe. The COVID Collection was broad in it's call for design, stating that it was created to ``\textit{inform decision-making on \ppe\ and medical device production, without stifling innovation}''. Interestingly, the community who submitted to this collection narrowed its focus to the production of three types of \ppe: masks, face shields, and ear-savers; \numourset\ of the \totalsubs\ total submissions (\oursetperc\%) fell into these three categories. The \numothersub\ ``other'' submissions focused on meeting a range of needs (e.g., ventilator parts, shoe covers, gowns, hand-less door openers, nasal swabs). 

\begin{figure}
\captionsetup{justification=centering}
\begin{subfigure}{.32\textwidth}
  \centering
  \includegraphics[height=1.4in]{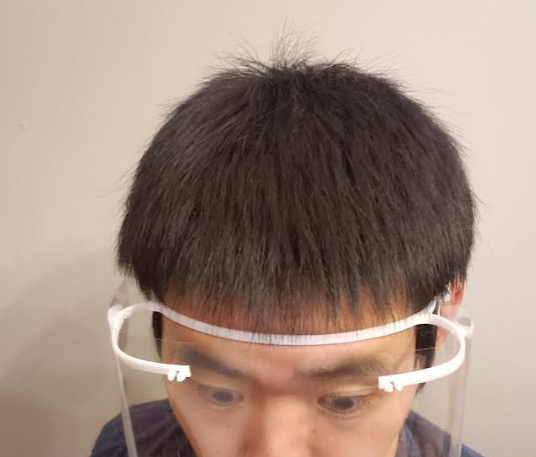}
  \caption{A face shield without protection from above (3DPX-013343)}
\end{subfigure}\hfill%
\begin{subfigure}{.32\textwidth}
  \centering
  \includegraphics[height=1.4in]{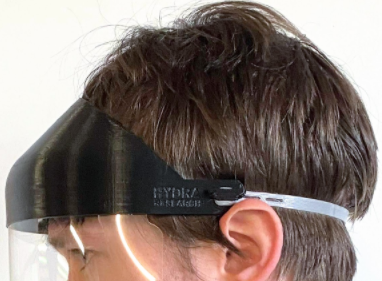}
  \caption{A face shield with protection from above (3DPX-013325)}
\end{subfigure}\hfill%
\begin{subfigure}{.32\textwidth}
  \centering
  \includegraphics[height=1.4in]{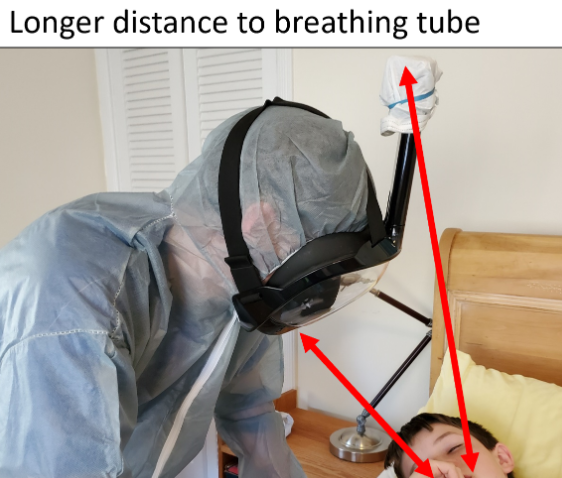}
  \caption{The scuba mask/face shield design (3DPX-013396)}
\end{subfigure}\hfill%
\caption{Examples of three types of face shields. The first two examples show the most common arcehtypes we found, those providing coverage from above (b) and those that do not (a). The third image (c) is an example of the ``scuba/snorkel'' designs that relied on a consumer face mask or snorkel mask that covers the whole face and air is breathed through the snorkel pipe. }
    \label{fig:shields}
    \Description{Three types of face shield. The first (sub-figure a) has a white 3d printed frame. It looks like a thin headband that wraps around the head, pressing against the wearer's forehead. this band has two hooks protruding from the front (over the forehead) on which a plastic, transparent sheet hangs to prevent liquid from reaching the face. Notable, there is a large gap that you can see between the 3D user's face and the transparency sheet, where droplets of liquid could fall.
    The second (sub-figure b) is the same as figure a except there is a small visor that starts at where the face shield touches the transparency sheet and slants up towards the top of the wearer's head. This visor prevents liquid from entering the space between the transparency sheet and the person's face.
    The third (sub-figure c) shows a person wearing a scuba mask. It has a tight seal around the user's entire face (chin to forehead) and is clear so the user can see out of it. There is a small tube that comes out of the sealed mask and points up towards the sky to allow air to reach the user under the sealed mask.}
\end{figure}

We further saw convergence of designs within these three overarching \ppe\ categories. Consider face shields. In our preliminary analysis of a random sample of face shields, the only common difference between the designs was protection from liquid droplets from above (\autoref{fig:shields}a and b). Besides this feature, face shields almost exclusively consisted of a 3D printed frame that braces against the forehead and a clear plastic sheet that attaches to the front of the frame to protect the face. The two most commonly remixed face shields (``3DVerkstan 3D printed face shield head band'' and ``DtM-v3.1 Face Shield \ppe'') followed this archetype. 
The convergence to only a few archetypes over a period of about a month is unusual. Generally, makers are espoused for their creativity and presentation of novel, innovative, even wild ideas. But those ideas were largely absent from the \nihEx. 

There was one notable design for a combined face shield-mask that starkly deviated from this norm: the ``Five-minute zero-print full-face snorkel mask with filter'' (3DPX-013396), shown in \autoref{fig:shields}c. It required no 3D printing and only  attachment of filtering material over the spout of a full-face, sealed, snorkel mask. There were 13 other scuba-mask-based designs that all used the same concept but used a 3D printed adapter to attach the filter material. Across our entire qualitative review, this was the only archetype that varied significantly from a design that was reviewed for clinical use before the rapid drop off in submissions in April. It is the exception that proves the rule.


\subsection{Safety}

The review system is the core component that distinguishes the \nihEx\ from any other maker repository. The process enforces clinical norms of safety and quality. Overall, the risks associated with different types of \ppe\ was the primary determinant in review status. More subtle details that contribute to safety quality were difficult to analyse because, to date, \notratedperc\% of designs have not been reviewed. However, some traits that we expect contributed to a design's safety-level could be found across the whole dataset. Though we are not experts in the safety of \ppe, we identified three relevant safety traits through our analysis: coverage, fit, and the presence of cleaning instructions. The safety criteria for masks and face shields differ, and so we discuss them separately below. Ear-savers, on the other hand pose little risk as an accessory to improve comfort, so we do not discuss their safety features. There are no examples of ear-savers with a ``warning'' usage rating status.

\subsubsection{Usage Ratings and Safety Results}
The \nihEx\ created four different usage ratings to classify entries based on the prototype's level of safety (\autoref{tab:approval}). The vast majority of entries (N=\numnotrated/\numourset, \notratedperc\%) had a ``prototype'' status, which is the default rating of submissions uploaded, indicating that the submission has not been fully reviewed. Though not an official rating, we did note that \percnotratedwnotes\% (N=\numnotratedwnotes/\numnotrated) of these submissions had received notes from the reviewer which indicates that these submissions were not acceptable given the level of documentation included in the submission. There were three statuses for designs that completed review, ``Clinical Use'', ``Community Use'', and ``Warning''. These categories dictate the level of trust reviewers had in the designs' safety and efficacy. Design affiliation with healthcare correlated with both likelihood of receiving reviewer attention ($\chi^2 = 11.4, p < .001$) and community or clinical ratings ($\chi^2 = 11.2, p < .001$). This may indicates that affiliated makers were sought out for review and were better suited to submit designs that reviewers viewed favorably (i.e., considered safe).

\begin{table}
\caption{The usage rating given to \ppe\ across the three main categories of face shield, mask, and ear-saver. The majority of designs received an unreviewed ``prototype'' status.}
\label{tab:approval}
\begin{tabular}{llll}
\toprule
\multicolumn{1}{l}{Rating}      & \multicolumn{1}{l}{Face Shield} & \multicolumn{1}{l}{Mask} & \multicolumn{1}{l}{Ear-Saver} \\ \midrule
Clinical use    & \clinicalfaceshield    & \clinicalmask     & \clinicalstrap    \\
Community use   & \communityfaceshield     & \communitymask    & \communitystrap     \\
Warning         & \warningfaceshield     & \warningmask     & \warningstrap     \\
Checked but no rating given & \notesfaceshield     & \notesmask    & \notesstrap \\
Prototype (not checked)     & \notcheckedfaceshield   & \notcheckedmask   & \notcheckedstrap                                  
\end{tabular}

\end{table}

\numclinicalrating\ entries (\clinicalratingperc\%) received clinical usage ratings, meaning the entry had been evaluated in a clinical setting and reviewers deemed appropriate for healthcare workers in contact with COVID-19 patients---their highest mark of safety. For example, the ``Stopgap Surgical Face Mask (SFM) Revision B'' (3DPX-014168) was evaluated in a clinical setting and was given a clinical usage rating. Others (N=\numcommunityrating/\numourset, \communityratingperc\%) received a community usage rating, meaning that the entry is suitable for workers in retail stores, law enforcement, and other community activities. \numwarningourset\ entries (\percwarningourset\%), both of which were masks, received a ``warning'' rating, indicating that the entry needed FDA approval or has design flaws that make it unsafe to use. Outside of our dataset of masks, face shields, and ear-savers, \percwarning\% of all submissions (N=\numwarning/\totalsubs) had a warning rating. A majority of these entries with warning status (N=15/\numwarning, 44.1\%) were ventilator parts. Many of these entries had notes from the author saying the entry had not been tested; for example, the author of the ``Ventilator Circuit Splitters - reinforced \& thicker walls'' (3DPX-013347) stated that they ``\textit{make no representations as to the safety of this device.}'' Other entries with the warning status included parts for other respiration devices and mask sanitizers. As we expected, classes of devices that pose more risk (e.g., masks, ventilator parts) received the more scrutiny, and devices that pose less risks (e.g., face shields, ear savers) received less scrutiny.

There were \numnotratedwnotes\ submissions that were not yet reviewed but had reviewer notes. The reviewer notes in a majority of these submissions (N=31/\numnotratedwnotes, 72.1\%) requested documentation, specifically best printing parameters or use instructions. Some reviewer notes (N=8/\numnotratedwnotes, 18.9\%) pointed out that the printing instructions and instructions for use were on external links and this documentation needed to be statically embedded in the submission to prevent modifications after review.
Other reviewer notes (N=4/\numnotratedwnotes, 9.3\%) requested that submissions be renamed so as not to imply incorrect usage and protection properties. For example, reviewers asked for ``respirator'' to be taken out of the title of the submission ``3D Printed Respirator Mask, 4 sizes, XSM, SM, M, L'' (3DPX-013948) because the term ``respirator'' is a medical term that implies a specific level of protection that this mask did not meet \cite{nih_faqs}. Two reviewer notes on masks requested testing information. For example, the ``The Unity Mask PRO''(3DPX-014364) listed that the mask met or exceeded National Institute for Occupational Safety and Health (NIOSH) N95 filtration criteria, but did not provide the test results. Based on these reviewer comments that makers' and reviewers' value of documentation were misaligned.

\subsubsection{Characteristics of Safe Designs}

Beyond the designs that were reviewed, we could only identify three characteristics of designs that we have a high confidence influence whether or not their reproduction is safe for clinical use: (1) mask sizing/fit, (2) face shield coverage, and (3) the presence of cleaning/disinfecting instructions). 

The main safety trait that varied across masks was the inclusion of different size 3D models. For masks, one size does not fit all; sizing is a key factor that influences fit and fit ensures safety. Facial features do not scale uniformly, so scaling model sizes is not a solution. Most masks
will not create a secure air seal on a diverse set of human faces. In these cases, contaminated air could enter through the gaps between the mask and face, rather than through the filter. Different sizes are needed to ensure that people of different ages and genders are protected. Only 43 (25\%) of the masks offered at least two sizes. In practice, users may find that different designs fit them better, but most wearers do not have the opportunity to print a range of masks and pick the best fit. The lack of sizing features in this data set shows that most makers were not considering this key safety feature in their design process. 


The main safety trait that varied across face shields was forehead coverage. Face shields at a minimum need to protect the front of the face (eyes, nose, and mouth) from liquid droplets, and almost all designs did so;  \numNOTfrontofface\ did not. However, several designs (\numcoverabove/\numfaceshield, \perccoverabove\% of face shields) also protected the wearer from liquid droplets from above by covering the forehead. Most designs either created a ``visor'' like piece to connect to the top of the mask or a closed gap so that there is no open space between where the frame touches the forehead and the clear sheet (see \autoref{fig:shields}). Forehead coverage may not be as critical as mask sizing for safety, but the additional feature demonstrates that many designers were considering increase safety when designing face shields.

A final piece of information that was critical to safe use of reusable \ppe\ in a pandemic was cleaning instructions. Cleaning instructions are necessary to ensure proper disinfection and safe reuse. Only \numcleaning\ (\perccleaning\%) and were included in instances of all three types of prototypes. Affiliation and usage rating were both correlated with presence of cleaning instructions. A $\chi^2$ test reveals that affiliation with a health care organization or a university correlated with the presence of cleaning instructions (healthcare: $\chi^2 = 5.0, p < .05$ ; university: $\chi^2 = 11.8, p < .001$), and the presence of cleaning instructions was correlated with a community or clinical usage rating ($\chi^2 = 33.3, p < .00000001)$.  Many cleaning protocols are based on common protocols for medical devices that are already in clinical use. Perhaps affiliated makers were more readily aware of these practices than unaffiliated makers. Notably, while cleaning instructions effected the review process, there was no submission field for including them explicitly.

Overall, there were only a few features that we could demonstrate impacted the safety rating of submissions. This may be because of how small the sample of reviewed designs is, making it difficult to identify common flaws in makers designs or characteristics of high-quality designs.

\section{Discussion}
The COVID-19 pandemic spurred one of the most widespread efforts of medical making to date. Makers sustained efforts to make \ppe\ with existing technical and human infrastructure \cite{Novak2020_covidremix}. The \nihEx\ COVID collection was meant to ``\textit{inform decision-making on \ppe\ and medical device production, without stifling innovation}'' \cite{nih_faqs}. Decision-making was supported by the formal review process, but reviewers were overwhelmed by the surge of designs. Further, the platform's goal of fostering innovation was undermined by regulated, inflexible structures which have been shown to create environments that can be less conducive for maker contributions \cite{khanapour2017framing}.
The decision to regulate the repository stems from the medical making community norm of safety and reliability. These prerogatives led medical makers to limit remixed designs, causing a rapid convergence of ideas. Hobbyist makers' contributions were subject to social and institutional structures of medical maker groups. In the context of medical making during a health crisis, we discuss how these material practices \cite{rosner_matcollab2012} inform future design of online repositories that support supporting innovation through peer production in high-risk domains.

\subsection{The Effects of Novel, Scarce Review}

\nih\ reviewers were inundated with hundreds of designs, and few designs were reviewed. Reviewer time is a scarce resource. Though the \nih\ provided some guidance for safely designing one type of \ppe, masks, it was not widely adopted by makers; few masks were approved and many lacked documentation. Further, the \nih\ provided makers little details on what the review process would consider. We suggest that this unclear reviewing process advantaged makers affiliated with healthcare institutions over lone-makers.


\subsubsection{The Benefits of Affiliation}
Makers who were affiliated with a university, the healthcare facility, or for-profit company were reviewed more positively because they tended to more successfully adhere to the \nih's value of safety. The designs affiliated makers contributed received more reviewer attention and higher usage ratings. While the \nihEx\ aspired to support collaboration, the collaborations that were positively reviewed tended to rely on institutional systems for accountability, reliability, and perceived value of contribution. Prior literature shows that maker communities are driven by social norms \cite{fox_2015_hacking, toombs_2015_enacting}. We contribute to this literature by demonstrating that social norms derived from clinical institutions advantage makers working in the healthcare domain.

Affiliated makers had better access to resources pools that aligned with the \nih's clinical expectations of safety. For instance, interviews with makers working in close proximity to healthcare workers helped affiliated makers evaluate \ppe\ usability \cite{lakshmi_covid}. Additionally, they could access rare medical expertise (e.g., infectious disease teams). Another study of COVID-19 maker communities showed that they struggle to curate and analyze scientific information in an evolving crisis plagued with widespread misinformation \cite{hofmann_onlineMakers}. This challenge further advantages affiliated makers with experience reading scientific literature (e.g., healthcare workers, university workers). In terms of material resources, universities and for-profit companies
often had resources like 3D printers, filament, and CAD and fabrication experts. Further, many healthcare and university workers had access to testing facilities, which explains the statistical correlations between these affiliations and presence of testing results . Finally, affiliated makers had access to teams of specialized experts when attempting faster iteration. Access to such resources is demonstrative of an institutional culture that supports and demands thorough documentation in order to decrease safety risks. It is to be expected that affiliated makers would bring these practices and the values they represent with them to the \nihEx.


\subsubsection{Reviewing Expert-Amateurs}
Individual or hobbyist makers may be unfamiliar with formal review processes, while affiliated makers are often accustomed to peer-review. 
Review is not common on maker repositories (e.g., Thingiverse), which support relatively unrestricted sharing of designs. Therefore, makers likely were unclear about expectations going in.
In our study, we saw that most submissions had the level of documentation we would expect to see on hobbyist repositories. However, the \nih\ reviewers needed documentation in order to replicate designs to review them. We see three plausible explanations for the discrepancy. First, makers may be defaulting to their usual practices and simply following the norms established on other repositories. Second, makers may lack an understanding of what details reviewers need, particularly early on when few accepted designs could be used as exemplars. Alcock \etal demonstrated with survey of Thingiverse that makers rarely provide enough documentation to support other makers in remixing their work \cite{alcock2016barriers}. It follows that they would continue to struggle to provide documentation on the \nihEx, but with greater consequence since it halts the review process.  Finally, makers may not value the review process as it excludes their contribution to the collaborative in clear terms \cite{morgan_peerlabor_2014}. Instead, we expect that many makers are using the \nihEx\ as a repository for COVID specific designs. The lack of documentation poses unique challenges for regulated maker repositories. How can makers be encouraged to document their work for review without discouraging makers with less resources or experience?

Lone makers seemed to struggle to create safe, documented designs, outside of making minor remixes of affiliated designs. Based on the designs they provided, unaffiliated makers appear to be unaware of the necessity of cleaning instructions or manufacturing instructions in the review process; understandably so, since cleaning designs is ``invisible'' \cite{morgan_peerlabor_2014} to makers outside of healthcare settings. However, the correlation between documentation and usage ratings indicates that this work is valued by reviewers. Therefore, novice medical makers can submit more effective designs if they are alerted to reviewers' values by encountering mandatory form fields asking for this safety critical information.

One could argue that  affiliated makers appear better prepared than lone makers to support healthcare settings---i.e., leave it to the professionals. But, this view erases the contributions of non-affiliated medical makers who provided critical support in the COVID-19 crisis \cite{hofmann_onlineMakers}. It also undermines the potential in harnessing low-skill craft alternatives in supplementing collective action beyond established practices \cite{lakshmi_covid}. Instead, we propose that the \nihEx\ naively positioned lone makers at the same level as large, professional, collaborative institutions. This forces makers to perform along standards set by larger institutions without those institutional resources. To do this, most makers tweaked accepted designs, rather than innovating in diverse ways. 

\subsection{Safety's Impacts on Design Diversity}

There were only a few distinct types of designs on the \nihEx\ and remixes with small incremental changes were frequent. However, the \nihEx's goal was not to ``\textit{stifle innovation}''; they wanted makers to contribute unique and novel solutions. We hypothesize that the norm of safety and the uncertainty of the review process encouraged individual makers to only make small changes when remixing.

We observed that when makers adapted others' designs, they limited their edits compared to the remixing behavior in other creative communities
; few makers made large, structural changes to designs. Affiliated makers contributed the designs that were remixed the most. These had also received community or clinical usage ratings. This data reveals a tendency of the community to follow norms set by perceived authorities including, these professionally affiliated groups and the \nih\ itself. However, as described, lone makers face greater challenges than institutions to document and test their designs. Instead, most makers likely work with scarce resources (e.g., one 3D printer model, few types of filament) and need to adapt designs to their emergent needs.

Lone makers had to two main strategies to meet the \nih's value of safety with their limited resources. One option was submitting a more innovative design without full documentation, and, therefore, receiving less reviewer attention or lower usage ratings. Otherwise, they posted derivative works of clinically reviewed designs. However small, there is inherent value in makers' incremental modifications of existing designs. Many modifications consisted of adjusting a design to allow faster manufacturing or printing with a different materials or devices. Unlike a comparable domain of open source software, Hudson \etal has shown that adapting designs to new fabrication techniques is no trivial feat \cite{hudson2016understanding}. In some cases, small changes (e.g., changing materials) may have unintended safety consequences, but, for the most part, these changes can help more people make a design. 

The remix behavior on the \nihEx\ emphasizes the affordances of physical materials and how they shape the designing and manufacturing process. That many remixes adjusted the materials or tools highlights an intention to interact with design materials as ``heterogeneous enactments'' over time rather than ``fixed forms'' \cite{rosner_matcollab2012}. We question measures of novelty as a degree of divergence from the original (e.g, \cite{cheliotisremix_2014}). The small remixes of designs modified material-determined affordances and expanded who could make well-rated designs. This helps more makers to aid in supplying \ppe during the pandemic. While we recognize these small and critical innovations, they are not inline with the \nih's expectations of makers. In a pandemic characterized by shortages in \ppe, using a range of materials or manufacturing methods and prioritizing speed along with safety are all valuable design goals. Therefore, seeking more unique design archetypes is a valuable goal.

Finally, makers may submit more divergent designs if they better understood higher level healthcare needs; one cannot innovate without first understanding stakeholder practices, values, and needs. The large number of masks, face shields, and ear-savers was linked to the media coverage of need for these types of \ppe. As specific designs got more positive coverage from their \nih\ reviews, makers were inclined to copy them. One solution to encourage novelty is to make the repository's high-level goals explicit. For example, while not directly connected to the NIH, the Department of Defense's America Makes initiative's design innovation contests were successful at producing novel \ppe\ that was submitted to the \nih\ exchange. Without clear calls to innovate, makers likely directed their efforts at the meeting the review process's afforded value of safety.  Rather than ``re-imagining what a face shield looks like'', they tweaked accepted designs.

\section{Design Recommendations}

The COVID-19 collection within the \nihEx\ has been an experiment in online sourcing of community medical device designs. In many ways, this tool is follows Lakshmi \etals recommendations to use  ``partially-open repositories'' to collect, review, and regulate medical makers' designs \cite{lakshmi2019point}. We understand, from the repository's statement, that it is intended to create an environment for purpose-driven collaboration with amateurs and experts \cite{Kuznetsov_2010_expertamateur}. Based on our findings, we conclude there are gaps to bridge in realizing this goal. We found that this iteration of the exchange was not able to review designs fast enough, and makers tended to submit risk averse designs rather than proposing novel and unique designs. While this trade-off may be inevitable, we expect that a balance between novelty and review could be struck with interface variations that support and reinforce community values like safety and innovation. We offer three suggestions.

\subsection{Clarify Reviewing Criteria}

A reviewing process is novel for maker repositories, and makers need clarifications to use it effectively. Confusion could be clarified by including reviewer comments in accepted designs. These comments are needed on incomplete and accepted designs so that makers can differentiate between the two. Makers would benefit from information about what made a design safe or effective. An additional way clarify requirements is to mark critical fields as mandatory for submission. 
However, the \nihEx\ may have limited requirements in an effort to not overwhelm makers. As a compromise, we propose that fields be marked as ``recommended for review''. This option would still highlight safety and documentation requirements without disenfranchising makers who prioritize sharing a design over achieving a usage rating. Reviewer time would be conserved since they could easily prioritize designs that have the details needed to replicate a design.

\subsection{Identify Remixes}

In it's current form, the \nihEx\ does not include a structure for annotating how a design was remixed. While this reduces burden of documenting changes, it also obfuscates ``material'' \cite{rosner_matcollab2012} distinctions between designs. Without a field to describe changes, it is up to makers to add this information in unrelated fields or to omit it entirely. 
Moreover, requiring these details could conserve reviewers' efforts by reviewing just the highlighted changes rather that re-reviewing derivative work.

\subsection{Motivate Innovation}

Designs on the \nihEx\ rarely diverged from a few common forms. Instead, it seems that safety was valued at the expense of innovation. To an extent we agree with how priorities were set; while innovation and creativity are important, safety is nonnegotiable. However, we expect two strategies can encourage innovation while ensuring safety. First, more diverse designs can be encouraged through explicit calls (e.g., ``build a better face shield''). Second, innovation can be rewarded along with safety, clarifying to makers that the two do not have to conflict. Safety was rewarded with clinical usage ratings. Similarly, we recommend that commendable innovation and creativity be noted in the review process. A ``uniqueness votes'' or ``tags'' system could encourage makers to explore new ideas. 
Reviewers could prioritize reviewing highly innovative designs over ones similar to reviewed designs. Since makers tend to modify designs with positive reviews this could have a snow ball effect where more makers remix designs that are increasingly divergent.

\subsection{Support Innovation}

To support the sharing of more creative, novel designs, collaboration interfaces for medical making must provide a structure to indicate the progress and/or intent of a design. All designs submitted to the \nihEx\ automatically received the ``prototype'' status, which put it in the queue for review. There was no way to designate a design as an ``seeking feedback'', ``not intended for production'', or ``ready for review.'' Consequently, we suspect this lack of affordance limited the scope of submitted designs to those that were close to current \ppe\ designs or reviewed designs on the \nihEx. Indeed, it is hard to determine the safety of a creative, novel design that is unlike previously reviewed designs. Posting such a design without indicating it is not ready to be manufactured can be unsafe, especially since new makers to the community may mistakenly view a designs affiliation with the \nih 's website as a sign of authority or approval. Introducing an ``in progress'' label to designs will encourage the sharing of more diverse ideas and seeking out feedback without risk of others adopting a design not ready for production.

\section{Limitations}
We recognize that our work was affected by our own experiences. Two of the authors were deeply involved in making \ppe\ this spring and contributed to designs submitted to the \nihEx. Though we have established relationships with the creators of the \nihEx, in this paper, we only draw from publicly available evidence. As researchers in computer science fields our recommendations focus on the design of tools, but we recognize that public policy determines what designs can be created and when and how they can be used. While we engage in wider discussions of policy, they are out of the scope of this paper. In particular, we have avoided making judgements about what makes designs safe or who should be doing this work. We leave such questions up to medical makers and suggest tools that could bolster these critical conversations. 


\section{Conclusion}
The \nihEx\ houses \totalsubs\ makers' designs for \ppe, the results of one of the most expansive efforts of medical making yet recorded. The forum was created to strike a balance between providing guidance through a formal review process and not stifling creativity. Our analysis of these \totalsubs\ reveals makers' misconceptions about the review process and criteria which lead to a rapid convergence of the design space. A few key designs created by university, for-profit company, and clinically affiliated makers received clinical usage ratings. Following submissions, particularly those made by unaffiliated makers, were derivatives of these designs. Often these submissions made small changes to optimize or increase flexibility of manufacturing. Overall, few designs were reviewed, and several of the designs that received reviewer attention were missing key pieces of information that prevented full review for clinical use.

In sum, our results suggest that affiliated makers received more positive ratings and more reviewer time than non-affiliated makers due to the knowledge and practices they bring from their clinical work. At the same time, many makers, particularly unaffiliated makers, often left out key pieces of information from their design submissions, leading to wasted review cycles. To make a more efficient and understandable review process without stifling maker creativity, we make three recommendations. First, prioritize unique designs for review to provide more examples of divergent and safe designs. Second, pose explicit requests to the community calling for diverse ideas and allow for makers to denote a design as ``seeking feedback'' so as to be clear that the mask is not ready for mass-manufacturing. Finally, establish clear metrics of safety. These changes aim to bridge the gap between the \nih's goals and unaffiliated makers' understanding of the review process and the values it implies.

\bibliographystyle{ACM-Reference-Format}
\bibliography{acmart.bib}

\end{document}